\documentclass[aps,preprint,groupedaddress]{revtex4-1}

\usepackage{amssymb}
\usepackage{amsmath}
\usepackage{graphicx}
\usepackage{enumerate}
\usepackage{mathtools}
\usepackage{subcaption}

\begin{document}


\title{Time's arrow and life's origin}


\author{Lucas Johns}
\email[]{ljohns@physics.ucsd.edu}
\affiliation{Department of Physics, University of California, San Diego, La Jolla, California 92093, USA}


\date{\today}

\begin{abstract}
The thermodynamic arrow-of-time problem is thought to be resolved by the observation that our universe initially was---and still is---far from equilibrium. The \textit{psychological} arrow-of-time problem is often attributed the same resolution, but the connection has not been thoroughly established. I argue that a compelling explanation of the psychological arrow requires an understanding of the physical conditions necessary for life to emerge from a prebiotic environment. A simple calculation illustrates how life-sustaining energy fluxes from the Sun and the Earth's interior bias the development of life in the direction of the arrow inherited from cosmic initial conditions.
\end{abstract}


\maketitle


The thermodynamic arrow of time---a symbol of irreversible macroscopic evolution from reversible microscopic laws---is typically traced back to the initial conditions of our universe \cite{davies1974, lebowitz1993, halliwell1996, price1996, zeh1999, carroll2008, ellis2013}. The arrow, it is said, points from a low-entropy, far-from-equilibrium past toward a higher-entropy, closer-to-equilibrium future. At first glance, the early universe may seem like anything but low entropy: cosmological measurements, after all, suggest that the universe prior to structure formation was exceedingly isotropic and homogeneous. But, the thinking goes, the high \textit{thermodynamic} entropy of the primordial plasma belies its low \textit{gravitational} entropy. This latter quantity admittedly has no agreed-upon definition, but the same point can be conveyed in more mundane terms: gravitational systems tend to clump, and the early universe was far from clumpy---ergo far from equilibrium. (For recent divergent approaches, see Refs.~\cite{maccone2009, jennings2010, barbour2014}.)

The psychological arrow of time, meanwhile, refers to the fact that we perceive time to be flowing unremittingly in one direction \cite{mlodinow2014, heinrich2014, rovelli2014}. Why, to put it one way, do we remember the past but not the future? To many a scientist, this sort of query sounds innocuous, if not downright silly, until they notice that the response ``because the future has not happened yet'' is not a claim evidenced by science. Yet it would be ill-advised to insist that science should be mute on this topic altogether, since in fact physics has much to say about time (as in \cite{isham1993, callender2011, anderson2012, balasubramanian2013} and all the literature cited above), and psychology much to say about our perception of it.

So what kind of answer \textit{is} evidenced by science? The most popular suggestion is that the psychological arrow follows from its thermodynamic counterpart, but the details of this suggestion are yet to be spelled out. Much more attention has been focused on the question of why the early universe had low entropy to begin with \cite{penrose1979, carroll2004} than on how exactly this initial state endowed human psychology with a perceptual arrow. And while the cosmological question is indeed a profound one, it is an insightful exercise to think carefully about the rest of the arrow's trajectory. With that mind, in this essay I will simply take initial conditions for granted and attempt to trace the more down-to-earth chain of logic that leads from cosmology to psychology.

One point warrants emphasis at the outset, to help us appreciate why any spelling-out is necessary to begin with. It is sometimes supposed that, since ``forward-living'' creatures such as ourselves observe the universe to obey the second law of thermodynamics, hypothetical reverse-living creatures would perforce observe a kind of reverse second law in which entropy always tends to decrease. This is not necessarily the case. Reverse-living scientists (by which I mean some hypothetical people whose arrow of time is the reverse of ours) would find the second law to apply just as well in their experiments as we do in ours, provided---and this is crucial---that we compare experiments in which the cosmic initial conditions are irrelevant.

A reverse-living scientist who, for instance, puts a collection of gas particles in the corner of an otherwise evacuated container and then releases them would witness the same homogenization of particle-number density that we would expect to see in our version of the experiment. A forward-living bystander, seeing the reverse-living scientist's experiment taking place, would observe an egregious violation of the second law, and yet the violation would be perfectly congruent with thermodynamics as a statistical theory. Violations of the second law are not impossible, but they are exponentially improbable---that is, unless there is fine tuning of initial conditions, in which case all bets are off. If the entropy of a system changes by $\Delta S$, then the number of microstates $\Omega_f$ encompassed by the final macrostate is related (assuming equiprobability) to the number of initial microstates $\Omega_i$ by $\Omega_f / \Omega_i = e^{\Delta S / k_B}$. A system appears fine-tuned at early times when $\Omega_f / \Omega_i \gg 1$, which is tantamount to large $\Delta S$. For reverse-living scientists to observe violations of the second law even in their own isolated experiments, their efforts to fine-tune their experimental setups ($\Omega_f \ll \Omega_i$) would have to be relentlessly outdone by even more severe fine tuning, in the opposite direction ($\Omega_f \gg \Omega_i$), concealed in the universe at large.

Put succinctly, any explanation of the psychological arrow should assume the fine tuning entailed by cosmic initial conditions, but no more than that. Therefore, if directionality in our psychology is a product of thermodynamics, it must be directly connected to the early universe being isotropic and homogeneous. But how?

The purpose of this essay is to offer an answer to that question, or at least an outline of one. I will suggest in particular that abiogenesis---the origin of life---is a crucial nexus linking the two arrows of time. Or to phrase the claim more provocatively, any compelling account of the psychological arrow must explain, outrageous though it may sound, why Earth evidently does not harbor organisms that live in reverse, moving (from our perspective) from death to birth. As argued below, the existence of such reverse life is disfavored by planetary thermodynamics, for reasons ultimately harking back to the initial state of the universe.

\begin{figure}
\includegraphics[width=.4\textwidth]{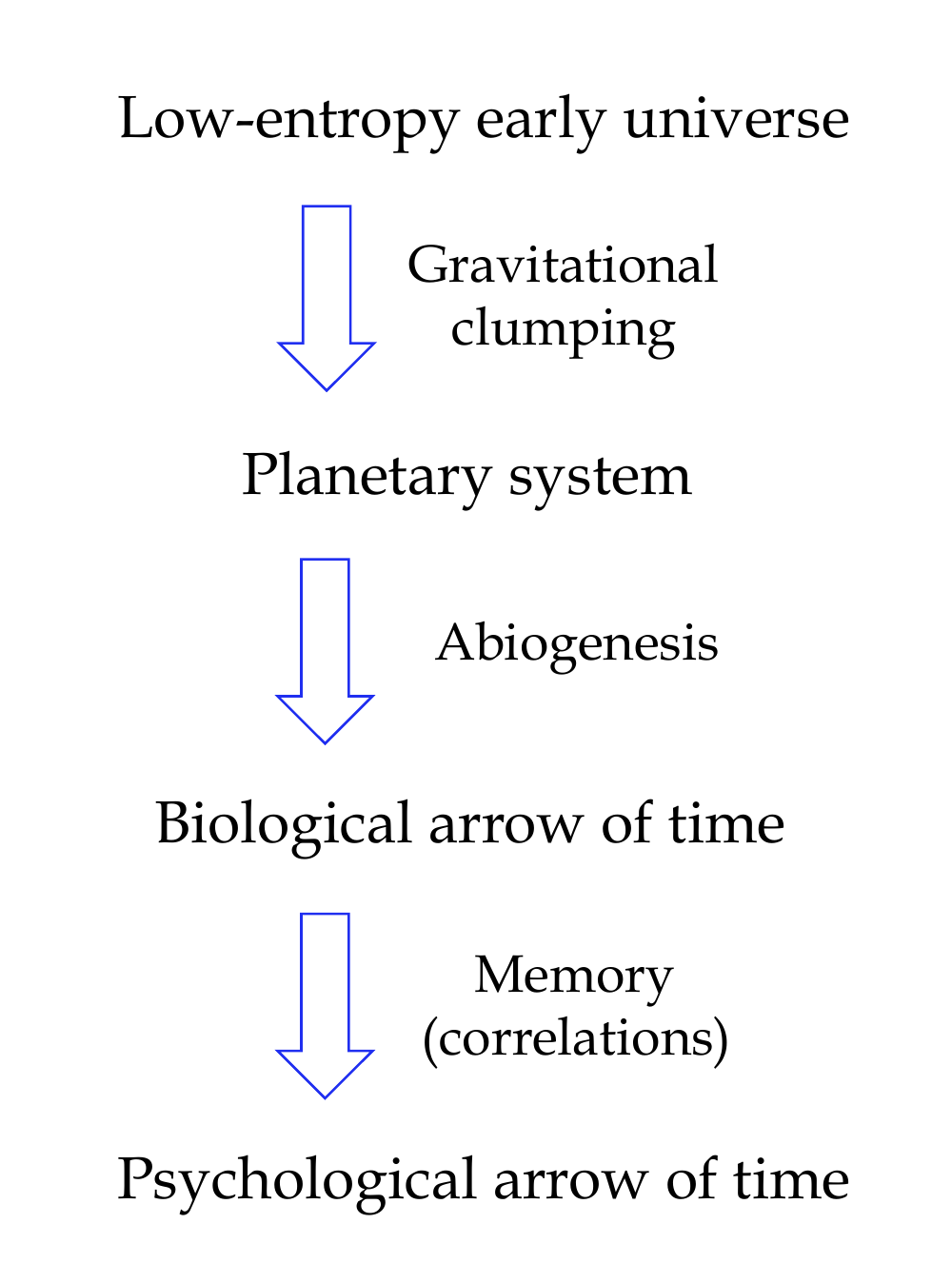}
\caption{The logical flow of the central argument, connecting the psychological arrow of time to the initial conditions of the universe. See text for explication.}  
\label{logic}
\end{figure}

The backbone of the argument is the following, going in reverse order (Fig.~\ref{logic}): Time perception is a biological function, and the psychological arrow is consequently inherited from what we might call the \textit{biological} arrow, which points in the direction of reproduction, metabolism, natural selection, and so on. For a group of organisms sharing a common ancestry, the biological arrow gets established once and for all at the origination of that lineage, with the arrow's direction being set by the physical conditions of the prebiotic environment. These conditions, in turn, must of course be consistent with the initial state of the universe, but the inheritance by biology (which is microscopic) of the thermodynamic arrow (which is cosmological) is not a foregone conclusion: as I have emphasized, it must be shown \textit{how} biology inherits the arrow. I will lay out a simplified example---an organism harvesting solar energy---that highlights the essential points while invoking only elementary thermodynamics.

Let us now go through the chain of logic in more detail, starting at the cosmic scale. A common trope in discussions of the thermodynamic arrow of time is to point to the Bekenstein--Hawking entropy \cite{hawking1971, bekenstein1973} of a black hole, $S_{BH} = k_B c^3 A /4 G \hbar$, as evidence that the entropy of the universe is increasing. The formula implies that the entropy of a single supermassive black hole can far exceed the thermodynamic entropy of all the matter and radiation in the visible universe. It is a great mystery what microstates $S_{BH}$ might be counting, but it is doubtful in any case that this entropy could be of direct relevance to human perception of time. The connection between cosmic initial conditions and the psychological arrow is slightly more roundabout: the isotropy and homogeneity of the early universe enabled gravitational clumping to proceed as the universe expanded and cooled, and the clumping in turn enabled the formation, ultimately, of stars and the kinds of celestial objects likely to host life. This development is indicated by the topmost arrow in Fig.~\ref{logic}.

Once life is established, so too is a collection of biological arrows. Organisms metabolize, maintain homeostasis, and reproduce, and all of these time-asymmetric behaviors point in the same direction. That these arrows are identical is not a coincidence, as all of the processes that sustain and propagate life require the utilization of free energy and thus necessarily align with the metabolic arrow. Reproduction then bestows this direction upon the arrow associated with natural selection. The latter is of course a prerequisite for producing organisms complex enough to even have a perception of time.

Since so little is known about how this faculty arises out of the perceiver's cognitive machinery, it is perhaps conceivable that the psychological arrow could nevertheless oppose the biological one, with the perceiver remembering death and anticipating birth. But the nature of memory counts heavily against this possibility. Regardless of the mechanism, memories of the external world must somehow correspond to correlations between organism and environment. The formation and storage of correlations in a retrievable form is almost certain to be metabolically costly in any organism (as suggested by a number of studies \cite{sartori2014, parrondo2015, bo2015}), and the direction of memory formation, hence the psychological arrow, is therefore expected to coincide with the other biological arrows.

But suppose, for the sake of argument, that this is not the case. Suppose that in fact an organism stores memories at no cost and, moreover, possesses a large enough storage capacity that Landauer's cost of erasure \cite{landauer1961} never needs to be paid. Perhaps, then, the psychological arrow can be unfastened from the metabolic one. This possibility is rendered profoundly unlikely by the fact that, in its distant past, such an organism must have had its memory apparatus prepared in a very special state. To take a toy-model example, if the apparatus consists of a collection of bits, each of which is realized (\`a la Szilard \cite{szilard1929}) by a particle in a box partitioned into two halves, then the state of each bit in the distant past must be known to the organism in order for the thermodynamic costs of memory formation to be evaded. This kind of blank-slate configuration of the apparatus betrays fine tuning. For the organism that remembers its youth, the fine tuning comes from its genetic material and embryonic development, with the start-up cost being paid by its progenitors. For it to remember its death instead, the fine tuning must come from the environment in which it perishes---a notion that beggars belief.

In short, the direction of our psychological arrow is an ancient bequest from our most primeval forerunners; this is indicated by the bottommost arrow of Fig.~\ref{logic}. The remaining question is why our ancestors had the biological arrow that they did. I propose that the answer is to be found in the physical conditions in which life originated, as indicated by the middle arrow of the figure.

Two forms of autotrophism are known to occur on Earth: photosynthesis, which relies on photons from the Sun, and chemosynthesis, which relies on the oxidation of inorganic compounds. Most theories of abiogenesis and early evolution are correspondingly predicated on the input of one of two energy sources \cite{fenchel2002}: solar radiation for surface-dwelling life \cite{rapf2016} or chemical fluxes for extremophiles (located, for instance, near hydrothermal vents \cite{martin2008}). The following simple calculation hints that it is these fluxes which bias the emergence of life toward forward-living organisms such as ourselves.

\begin{figure}
\includegraphics[width=.80\textwidth]{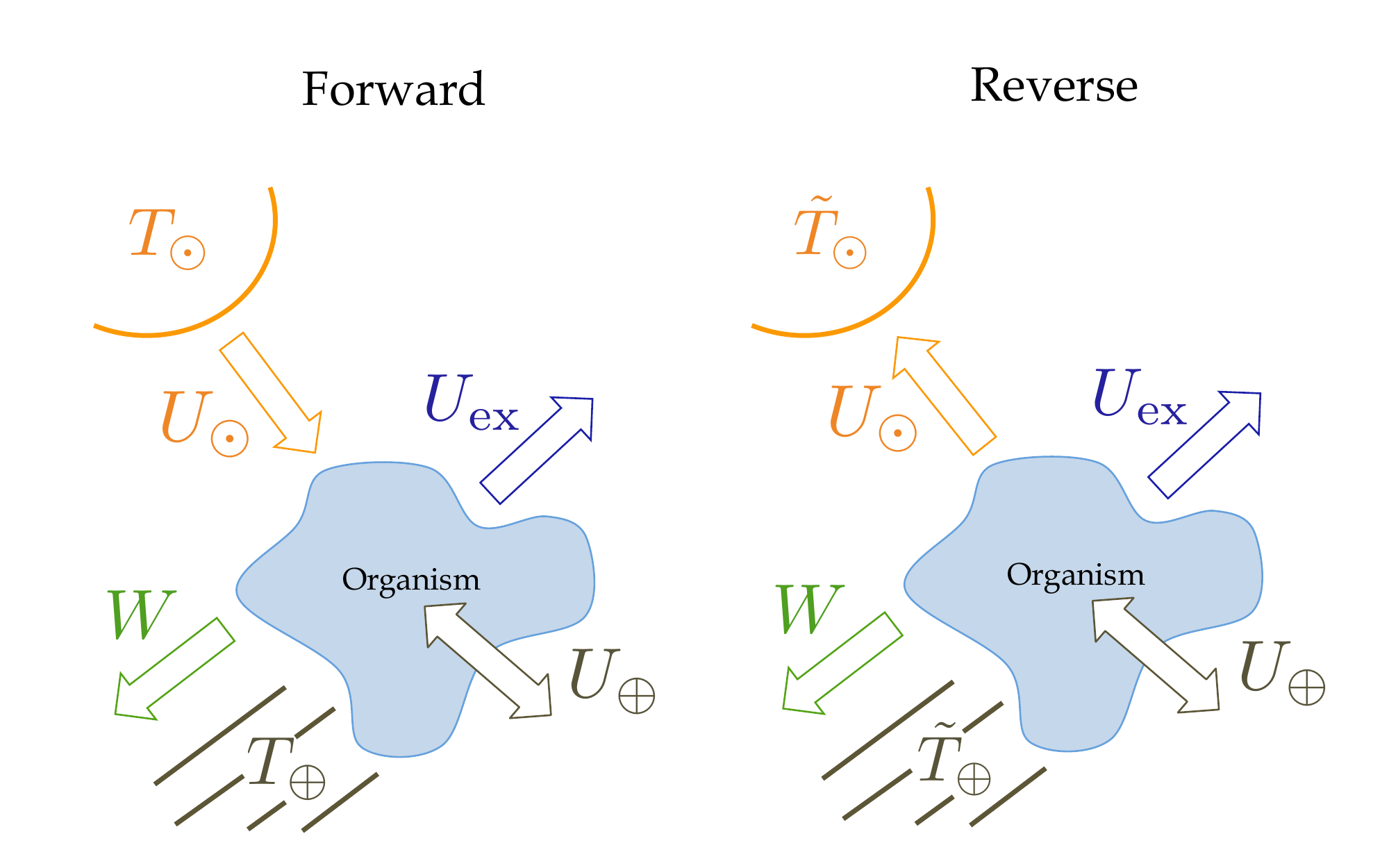}
\caption{Energy flows for a forward-living organism subsisting on solar radiation (left) and a reverse-living organism attempting to subsist in the time-reversed environmental conditions (right). The crucial difference is whether solar energy $U_\odot$ is absorbed by or extracted from the organism. The meaning of the tildes in $\tilde{T}_\odot$ and $\tilde{T}_\oplus$ is discussed in the text.}  
\label{solar}
\end{figure}

The left panel of Fig.~\ref{solar} depicts the utilization, over some period of time, of solar energy $U_\odot$ by an organism, which over the same period excretes an energy $U_\textrm{ex}$, does work $W$ on its surroundings, and loses heat $U_\oplus$ to the Earth, which is taken to be at temperature $T_\oplus$. If the system is in a steady state, then $U_\odot - U_\textrm{ex} - W - U_\oplus = 0$. The second law of thermodynamics requires the relation $\Delta S_\textrm{org} \ge - \Delta S_\textrm{env}$ between the organism's entropy change and the environment's. Breaking down $\Delta S_\textrm{env}$ into its individual parts gives
\begin{equation}
\Delta S_\textrm{org} \ge \Delta S_\odot + \Delta S_W - \Delta S_\textrm{ex} - \Delta S_\oplus, \label{entropy}
\end{equation}
where the terms on the right-hand side are the entropy changes associated with $U_\odot$, $W$, $U_\textrm{ex}$, and $U_\oplus$, respectively.

Let the organism be at the ambient temperature $T_\oplus$ (or perhaps at $T \neq T_\oplus$, but shielded from heat exchange with the Earth), so that $U_\oplus = \Delta S_\oplus = 0$. The solar energy can then be used entirely for two purposes: doing work and preventing the growth of (or reducing) internal entropy. The relative expenditures on these functions will depend on the organism's physical make-up. If it seeks to maximize $W = U_\odot - U_\textrm{ex}$, then it is only limited by the efficiency with which it can convert solar energy into work. If instead it seeks to reduce its internal entropy as much as possible, then it is only limited by how effectively it can degrade solar energy (that is, how large it can make $\Delta S_\textrm{ex}$), echoing fundamental results from Refs.~\cite{england2013, perunov2016}. For example, if the organism emits waste only in the form of long-wavelength blackbody radiation at temperature $T_\textrm{ex}$, then the most it can reduce its internal entropy is
\begin{equation}
\Delta S_\textrm{org} = - \frac{4}{3} U_\odot \left( \frac{1}{T_\textrm{ex}} - \frac{1}{T_\odot} \right). \label{radiate}
\end{equation}
In practice the temperature of the sky will limit how low $T_\textrm{ex}$ can be, and inevitably there will be restrictions based on the organism's constituent biochemistry. But in any case, so long as the organism possesses adequate internal machinery, $W$ and $\Delta S_\textrm{org}$ can be made arbitrarily large and small, respectively, simply by capturing more photons from the Sun.

The idea now is to consider how productive an organism can be when this thermodynamic model is time-reversed. If it proves not to be viable in the time-reversed model for any organism to do work on its surroundings or to prevent its internal entropy from growing, then we have identified an explanation for the emergence of a forward-directed biological arrow.

To this end, let us consider the right panel of Fig.~\ref{solar}, which shows the time reversal of the left panel. Note that the arrows associated with $W$ and $U_\textrm{ex}$ should \textit{not} be flipped, because the question we are interested in is whether life can exist if the relevant features of the \textit{environment} have been time-reversed. The hypothetical organism must do work and excrete energy just like its forward-living analog, lest it lack the fundamental traits of being alive. Tildes are now affixed to the temperatures of the Sun and Earth, to denote that these objects act as time-reversed blackbodies. Rather than emitting photons in all directions, they now \textit{absorb} photons from all directions. Radiative equilibrium between two celestial objects at $\tilde{T}_\odot$ and $\tilde{T}_\oplus$ therefore implies a violation of the second law of thermodynamics that precisely mirrors the increase in entropy generated by radiative equilibrium between objects at $T_\odot$ and $T_\oplus$.

In other words, we are facing the physically appalling notion of photons getting pulled out from every point along a line of sight from the Sun. The scenario offends intuition, but the offensiveness is in exact proportion to the asymmetry inherent in the radiative arrow of time. The strangeness is in fact a clue that we are on the right track.

Again assuming a steady state, the work done by the organism is $W = U_\oplus - U_\odot - U_\textrm{ex}$. It is immediately clear that the organism will encounter energetic difficulties, as it must not only be able to harvest ambient heat energy but must be able to do so prodigiously enough that it can fuel its biological activities \textit{and} offset the demand from $\tilde{T}_\odot$. The task would appear to be insurmountable, and entropy considerations only doom the prospects further. The violation of the second law that goes along with the processing of some of $U_\oplus$ into $U_\odot$ reflects the existence of fine tuning in the degrees of freedom comprising $U_\oplus$. Some of the energy, to put it one way, is earmarked for emission toward the Sun. The organism, however, is clueless as to which degrees of freedom are fine-tuned in this way. Its inability to trust that its fuel source will obey the second law is fatal.

To be clear, straightforward application of the second law as in Eq.~\eqref{entropy} does not, on its own, preclude the thermodynamic viability of reverse life. The requirement in the reverse case is
\begin{equation}
\Delta S_\textrm{org} \ge - \Delta S_\odot + \Delta S_W - \Delta S_\textrm{ex} + \Delta S_\oplus, \label{entropy2}
\end{equation}
and it is evident that here (as in the forward direction) the second law can be satisfied, while at the same time maintaining $\Delta S_\textrm{org} \lesssim 0$, if a sufficiently large $\Delta S_\textrm{ex}$ is generated. The problem vis-\`a-vis entropy is that Eq.~\eqref{entropy2} does not reflect the hidden fine tuning connected to cosmic initial conditions. Returning to the example where the organism excretes energy in the form of radiation, Eq.~\eqref{radiate} naively becomes, in reverse,
\begin{equation}
\Delta S_\textrm{org} = - \frac{4}{3} U_\odot \left( \frac{1}{T_\textrm{ex}} + \frac{1}{T_\odot} \right).
\end{equation}
But this equation is misleading, because in reality the fine tuning undermines the use of blackbody expressions for entropy. There may be a way to compute the entropies involved in the reverse scenario such that fine tuning is accounted for, but the question need not concern us here because any such formulae are biologically irrelevant: information must be known in order to be useful \cite{sagawa2012, deffner2013, goold2016}.

The thermodynamics near hydrothermal vents is different in detail but similar in essence, with the solar flux replaced by thermal and chemical gradients sustained by effluence from the fissured ocean floor \cite{russell1997, simoncini2010}. Just as $\tilde{T}_\odot$ sabotages the efforts of reverse life to form under the conditions of Fig.~\ref{solar}, so too do the time-reversed fluxes in the vent system preempt reverse life, by sapping energy from the environment and its would-be inhabitants. Fluxes from the Sun and the Earth's interior therefore bias the origination and development of life in the forward direction, as symbolized by the middle arrow in Fig.~\ref{logic}.

The thermodynamics of life is thus a crucial link connecting the psychological arrow of time to the initial conditions of the universe. Inasmuch as life necessarily harnesses time-asymmetric fluxes, the direction of the arrow appears to be a fait accompli from the earliest moments in cosmic history.

\begin{acknowledgments}
I would like to thank Eve Armstrong, Pat Diamond, Tucker Elleflot, and George Fuller for delightful conversations. This work was supported by NSF Grant No. PHY-1614864.
\end{acknowledgments}

\bibliography{all_papers}

\end{document}